\documentstyle[emulateapj,psfig]{article}
\begin{document}    
\newcommand{\be}{\begin{equation}}
\newcommand{\ba}{\begin{eqnarray}}
\newcommand{\ee}{\end{equation}}
\newcommand{\ea}{\end{eqnarray}}  
       
\title{Linking the Metallicity Distribution of Galactic Halo Stars
	to the Enrichment History of the Universe}

\author{Evan Scannapieco\altaffilmark{1} \& Tom Broadhurst\altaffilmark{2}}
\altaffiltext{1}
{Department of Astronomy, University of California, Berkeley, CA  94720}
\altaffiltext{2}
{European Southern Observatory, Garching bei   M\"{u}nchen, Germany}

\begin{abstract}

We compare the metallicity distribution of Galactic Halo stars with 3D
realizations of hierarchical galaxy formation.  Outflows from dwarf
galaxies enrich the intergalactic medium inhomogeneously, at a rate
depending on the local galaxy density. Consequently, the first stars
created in small early-forming galaxies are less metal-rich that the
first stars formed in more massive galaxies which typically form
later. As most halo stars are likely to originate in accreted dwarfs,
while disk stars formed out of outflow-enriched gas, this scenario
naturally generates a ``metallicity floor'' for old disk stars, which
we find to be roughly coincident with the higher end of our predicted
metallicity distribution of halo stars, in agreement with
observations.  The broad and centrally peaked distribution of halo
star metallicities is well reproduced in our models, with a natural
dispersion depending on the exact accretion history.  Our modeling
includes the important ``baryonic stripping'' effect of early
outflows, which brush away the tenuously held gas in neighboring
pre-virialized density perturbations. This stripping process does not
significantly modify the predicted shape of the halo star metal
distribution but inhibits star-formation and hence the number of
accreted stars, helping to reconcile our model with the observed total
Galactic halo luminosity and the lack of low-luminosity local
dwarf galaxies relative to N-body predictions.

\end{abstract}


\section{Introduction}

There is no completely successful model for the metal distribution of
stars in our Galaxy. The simple `one-zone' model for the disk, in which
enrichment increases with time, is a poor match to the data, (van den
Bergh 1962; Schmidt 1963) generating a broad and weakly peaked
distribution. Better agreement is found by extending the 
disk-formation time with a declining gas inflow rate (Larson 1974;
Pagel 1989) but this cannot explain the sharp cut-off in the
number of G-dwarfs with metallicities below [Fe/H] $\sim -1$.  
In order to explain the full distribution,
variable infall must be complemented by
an {\em ad hoc} initial production spike or ``prompt initial
enrichment'' (e.g., Truran \& Cameron 1971; Ostriker \& Thuan 1975;
K\"oppen \& Arimoto 1990).

In the Galactic Halo, a very different enrichment history is required.
Here the metallicity distribution is broader with a peak shifted to a
lower value of [Fe/H] $\sim -1.5$ and is over-abundant in $\alpha$
elements compared to the disk, implicating predominantly supernova Type-II
enrichment (e.g., Gilmore \& Wyse 1998).  
No sharp cutoff is found but rather a long tail extending
to low values of [Fe/H] $= -3$ and below (Ryan \& Norris 1991). Unlike
the disk, this metal distribution is generally understood as a sum of
individual dwarf galaxies accreted by the Galaxy. This idea was first
suggested by Searle \& Zinn (1978) and survives today in hierarchical
models, such as the Cold Dark Matter (CDM) model (e.g., Kauffman,
White, \& Guiderdoni 1993; Cole et al.\ 1994).  Observationally, this
scenario is strongly supported by tidally disturbed objects, in
particular the Sagittarius dwarf galaxy and its associated stream of
Carbon stars (Ibata, Gilmore, \& Irwin 1994; Ibata et al.\ 2000), by
concentrations in the velocity and spatial distribution of halo field
stars (Preston, Beers, \& Shectman, 1994; Carney et al.\ 1996, Helmi
\& White 1998), and the large spread in the ages of both the halo
field and halo cluster stars (Laird \& Sneden 1996; Sarajedini,
Chaboyer, \& Demarque 1997).  For a different point of view, see
however, Oey 2000.

Despite the growing consensus that the Halo was assembled
hierarchically, a realistic description of this process has not been
developed.  Argast et al.\ (2000) consider a 3D model of interstellar
medium evolution  that roughly matches the observed properties
of the Halo but is unconnected to the accretion of dwarf objects (see
also Tsujimoto, Shigeyama, \& Yoshii 1999).  C\^ote et al.\ (2000)
adopt an empirical approach, constructing the metal distribution from
the observed metallicity-luminosity relation of low-mass galaxies.
This model, while roughly successful, is at odds with the observed
paucity of local dwarf galaxies (Klypin et al.\ 1999; Moore et al.\
1999) and leaves open the issue of the origin of metals in the dwarf
objects themselves.  Finally, Hernandez \& Ferrara (2000) examined the lowest
metallicity halo stars in terms of the mass-weighted 
average intergalactic medium (IGM) metallicity for each dwarf object 
formed, an approach which is most reliable for lower metallicity values.

In this letter we offer a holistic 3D picture in which the
physics behind the luminosity-metallicity relation, the cause of
``prompt initial enrichment'' of the Galactic disk, and the mechanism
that prevents an excess of local group dwarfs all stem from the same
process.  Galaxy outflows enrich the IGM slowly and inhomogeneously
over time, so that the first stars formed in small early galaxies are
more metal poor than the first stars formed in larger, later-forming
objects.  
As pre-virialized overdensities that would otherwise form
galaxies are especially susceptible to disruption by winds from
neighboring objects (Scannapieco, Ferrara, \& Broadhurst 2000;
Scannapieco \& Broadhurst 2001, hereafter SB; Scannapieco, Thacker \&
Davis 2000) galaxy outflows preferentially suppress the formation of
late-forming dwarf galaxies, reducing the total number of local
group dwarfs and Galactic Halo stars.

The structure of this work is as follows.  In \S2 we describe our
semi-analytical simulations of galaxy formation and metal dispersal,
and our criteria for selecting MW-like objects.  In \S3 we
present the results of these simulations in relation to the
Halo, with conclusions in \S4.

\section{Models of the Milky-Way Stellar Distribution}

\subsection{Simulation of Galaxy Formation}

Using the semi-analytical model described in detail in SB, we construct
a (12 Mpc/$h$s)$^3$ cubic comoving volume, where $h$ is the Hubble
constant in units of 100 km/s/Mpc, and track the collapse of dark
mater halos, cooling of gas, and formation of galaxy outflows.  A
(512$)^3$ linear density field is constructed from a standard fit to
the galaxy power spectrum and convolved with spherical ``top-hat''
filters corresponding to ten different mass-scales, spaced in equal
logarithmic intervals from $1.1 \times 10^8 M_\odot$ to $5.3 \times
10^{12} M_\odot$ and covering the interesting range from the most massive
galaxies to objects that lie close to the lower limit set by
photo-ionization and molecular cooling (SB).  Using the elliptical
model for collapse developed in Sheth, Mo, \& Tormen (1999) we search
through these fields, identifying the collapse redshifts and locations
of dark-matter halos.

Additional modeling is required to reproduce the formation of
galaxies.  The two important issues in this case are the delay between
halo collapse and the cooling of the virialized gas into stars, and the
ejection of material into the IGM by the outflows generated in the
initial star-formation episode.

To account for the delay between collapse and star-formation, we use
the standard inside-out collapse model, calibrated against numerical
simulations (White \& Frenk 1991; Somerville 1997).  The cooling
gas is modeled as an initially isothermal sphere $(\rho(r) \propto r^{-2})$
at the virial temperature $T$ and with a constant
metallicity, $Z$ calculated as described below.  All the gas within a
radius $r_{\rm cool}$ cools instantaneously according to a $\rho^2
\Lambda(T,Z)$ radiation law, where $\Lambda$ is the radiative cooling
function, and the gas outside this radius stays at the virial
temperature of the halo, with $r_{\rm cool}$ moving outward with time.
When the total mass contained within $r_{\rm cool}$ equals the
object's baryonic mass, we consider a new galaxy to be formed.

Once an object has collapsed and cooled into a galaxy, we assume that
a fixed fraction of the stellar material $\epsilon_{\rm sf}$ is
converted into stars in an initial burst of star-formation and that an
outflow develops powered by supernovae and stellar winds. We estimate
that one massive star forms for
every $100 M_\odot$ of stars, with an associated 
energy output of $1 \times 10^{51}$ ergs in stellar winds during its
lifetime and an additional $1 \times 10^{51}$ ergs when it explodes. 
A fraction $\epsilon_{\rm wind}$ of this energy is
channeled into a galaxy outflow, ejecting gas into the IGM.

The outflows are modeled as spherical shells that are driven by
internal pressure and decelerated by gravitational breaking by the
halo (see SB), both estimated in the thin shell approximation
(Ostriker \& McKee 1988).  These shells expand into the Hubble flow,
sweeping up the IGM and losing only a small fraction $f_m = 0.1$ to
the interior.  Their evolution is then completely determined by the
assumed luminosity history of the blast wave, and we follow
Tegmark, Silk, \& Evrard (1993) in considering only four
contributions: energy injection from supernovae, heating from
collisions between the shell and the IGM, cooling by Compton drag
against the microwave background, and bremsstrahlung radiation and
other two-body interactions.

The metallicity of each collapsing object is volume averaged over the
contributions of each of the expanding blast waves passing within the
collapse radius $r_{\rm coll}$, adopting an average yield from each
supernovae of 2 $M_\odot$ in metals (e.g., Nagataki \& Sato 1997).
Each object is assigned a mass in metals $M_Z$, taken to be
zero initially and modified by each outflow passing within $r_{\rm
coll}$.  For each such occurrence,  $M_Z$ is updated to
\be
M_Z \longrightarrow M_Z   + 
\frac{V_{\rm overlap}}{\frac{4 \pi}{3} r_{\rm coll}^3}
M^{\rm blast}_Z 
\label{eq:met}
\ee
where $V_{\rm overlap}$ is the volume of intersection 
between the outflow and the collapsing sphere.
By dividing this mass by the total baryonic mass of the
galaxy we can compute the initial metallicity of the object.

Our outflow models incorporate the ``baryonic
stripping'' mechanism identified in Scannapieco, Ferrara, \&
Broadhurst (2000), whereby outflows from early galaxies strip the gas
out of nearby overdense regions that would have otherwise become
relatively low-mass galaxies.  Stripping occurs whenever a shock moves
through the center of a pre-virialized halo with sufficient momentum
to accelerate the gas to the escape velocity, $v_e$, and thus we
exclude all objects for which
\be
	f \, M_s \, v_s \geq M_c \, v_e,	
\label{eq:strip}
\ee
where $f $ is the solid angle of the shell subtended by the
collapsing halo, $M_s$ is the IGM mass swept up by the
shock, $v_s$ the velocity of the shock, and $M_p$ is the baryonic mass of
the pre-virialized halo.

\subsection{Identification of Milky-Way Type Galaxies and Construction
	of Stellar Distributions}

Having developed a model of galaxy formation,
we now select MW-like galaxies from this
simulation. Mass estimates of our Galaxy from line-of-sight
satellite velocities place a lower limit of  $\approx 10^{12}
M_\odot$ (e.g., Kulessa \& Lynden-Bell 1992; Binney \& Merrifield
1998) within $\approx 50$kpc/h.  For the elliptical collapse
model employed in our calculations the total mass is greater
however, corresponding to the larger virial radius $r_{200}$, where
the average interior overdensity is 200. N-body comparisons between circular
velocity and total halo mass within $r_{200}$ 
suggest values $\sim 2.5 \times 10^{12} M_\odot$ (e.g., Steinmetz \&
Navarro 1999).  We adopt this approximately and select halos of $1.6
\times 10^{12} M_\odot$, the closest mass-scale in our
simulations and a value similar to $2.0 \times 10^{12} M_\odot$
used in Hernandez and Ferrara (2000).

A second important property of the Milky Way is that it has not had a
major merging event. The morphology-density relation (e.g., Abell
1958; Balland, Silk \& Schaeffer 1998), as well as numerical studies
of galaxy formation (e.g., Aguilar \& White 1985; Hernquist 1993)
suggest that the spherical components of galaxies form by major mergers.
As the bulge of our Galaxy represents
$\lesssim 10 \%$ of the total mass (e.g., Rohlfs \& Kreitschmann 1988;
Dehnen \& Binney 1998) we therefore  exclude all objects with more than one
progenitor on the two mass-scales below that of the MW ($4.8 \times
10^{11} M_\odot$ and $1.4 \times 10^{11} M_\odot$).

The final constraint on the identification of MW candidates is the age
of the assembled product, which should be relatively large to match
the age of the MW disk.  This is important as the
IGM is formed over a protracted period, and thus a galaxy that formed
much earlier or later is likely to have a different metallicity.  Note
that the time at which the collapsed gas within the $1.6 \times
10^{12} M_\odot$ object cools is associated with the formation of the
Galactic disk only. A reliable measure of this age is given by the
lower-limiting  luminosity of white dwarfs in the disk,
 which cool with time (Schmidt
1959). Improvements in both observations and our understanding
of  white dwarf structure
 have led to a consensus of approximately 10 Gyr for the
age of the disk (e.g., Winget et al.\ 1987; Leggit, Ruiz, \& Begeron
1998; Knox, Hawkins, \& Hambly 1999).  In our simulations we allow all
objects that formed 10 $\pm 3$ Gyrs ago to be considered as MW
candidates.

Within each MW-like galaxy, we use the full
merger history to divide its stellar population into disk,
bulge, and halo stars.  The first generation of disk
stars are tagged as those formed in the MW-like object
itself, while the other two populations are formed within
progenitor objects.  Whether progenitor stars are absorbed
into the Galactic Halo or instead become part of the bulge
depends on their initial trajectories.  Here we apply the rough rule
of thumb that the bulge was formed out of objects contained in the
inner $10 \%$ of the Galactic dark-matter halo, 
while the stars in objects in the remainder of the collapsed region
were accreted but not absorbed into the central bulge.  As the orbits
of these objects are much different than that of the main
disk component, they are naturally associated with the Halo (e.g., Bekki \&
Chiba 2000).

Note that the relatively early formation times of the MW candidates,
corresponding to even earlier virialization times for their
dark matter halos, mean that the majority of the
the progenitor galaxies used to construct the halo distribution
are formed at relatively high redshifts $gtrsim$ 2.  These
early times, coupled with a large minimum progenitor mass set
by photo-ionization and molecular cooling concerns, insures that
our semi-analytical treatment can be trusted
to construct an accurate merging history for each candidate.

Note also that our simulations only provide the metallicity of the first
generation of stars to form in each object, and no subsequent
star-formation is considered.  Fortunately all accreted objects are
sufficiently small ($\lesssim 5 \times 10^{10} M_\odot$) that second
generation star-formation is quashed because much of the gas is
expelled. This is consistent with observations, which suggest 
that only a small fraction of stars in the halo were enriched by
Type IA supernovae (Gilmore and Wyse 1998).
Hence the observed halo stellar metal distribution can be
reliably compared to our simulations.  Because our
filtering quantizes the galaxy mass distribution, we incorporate a
small smoothing of the metallicity of each galaxy, modeling it with
a mean metallicity as calculated from Eq.\ (\ref{eq:met}) and a
dispersion of $\sigma$([Fe/H]) = 0.3 independent the mass and
mean metallicity of the dwarf (C\^ote et al.\ 2000), a procedure that
introduces no significant bias and is more realistic than
adopting a single metallicity for all the stars in a given object.

\section{Results}
With these prescriptions we generate and identify model
MW-like galaxies, which are massive, old, and have suffered 
at most a merger with less than 10\% of their total mass. 
  We restrict our attention to the currently favored
$\Lambda$CDM cosmology and fix $\Omega_0$ = 0.35, $\Omega_\Lambda$ =
0.65, $\Omega_b = 0.06$, $\sigma_8 = 0.87$, $\Gamma = 0.18$, and
$h=0.65$, where $\Omega_0$, $\Omega_\Lambda$, and $\Omega_b$ are the
total matter, vacuum, and baryonic densities in units of the critical
density, $\sigma_8$ is the amplitude of mass fluctuations on the 8
$h^{-1}$ Mpc scale, and $\Gamma$ is the CDM ``Shape Parameter.''

In our fiducial outflow model we adopt the same parameters used in SB
and Scannapieco, Thacker, \& Davis (2000) and take $\epsilon_{\rm sf}
= 0.1$ and $\epsilon_{\rm wind} = 0.1.$ For comparison we also
consider a ``low-energy outflow'' simulation with $\epsilon_{\rm
sf} = 0.1$ and $\epsilon_{\rm wind} = 0.05$, and a simulation in which
we again set $\epsilon_{\rm sf} = 0.1$ and $\epsilon_{\rm wind} = 0.1$
but do not apply the baryonic stripping criteria, Eq.\
(\ref{eq:strip}).  In the fiducial and low-energy outflow models,
roughly two thirds of the objects in the $10^9 M_\odot$ to $10^{11}
M_\odot$ mass range are suppressed by stripping, helping to
reconcile CDM predictions with the observed lack of visible
local group dwarf galaxies.  In all cases, the average initial
metallicity of galaxies ranges from approximately [Fe/H] = -2.0 to
[Fe/H] = -0.6, increasing roughly linearly with the log of the mass.
Further details are presented in SB.

The metallicity distribution of halo stars in the five MW-like
galaxies identified are shown in Figure 1, along with the observations
of MW Halo stars by Ryan and Norris (1991).  As each simulation shares
the same random seed, these model distributions are directly comparable
across runs.  In all cases, we see that the initial disk metallicity 
coincides with
the high-metallicity end of the halo star metal distribution,
mimicking the behavior of the MW.  While the
low-energy outflow and fiducial models produce similar results, reducing
$\epsilon_{\rm wind} = 0.05$ slightly lowers $Z$
values, more in line with the observations but well within the
uncertainties of our approach.  Note that the metallicities in our
models can be shifted linearly by altering the assumed stellar yields.

Exclusion of baryonic stripping has a larger effect, producing a
sharper peak and a factor of $\sim$ 3-5 more halo stars.
While the overall shape of the distribution is reasonable,
this model
suffers from the same excess of low-mass dwarf galaxies as the
empirical model suggested in C\^ote et al.\ (2000). The models
that include stripping via Eq.\ (2), on the other hand,
 can be more easily reconciled with the observed
luminosity of the Halo: $L_{\rm V} \approx 5 \times 10^7
L_\odot$ (Binney \& Merrifield 1998) corresponding to
a total stellar mass of $\sim 5 \times 10^8 M_\odot$
for a Scalo (1986) initial mass function
and a $\sim 10^{10}$ year old stellar population.
Note that as more winds are formed at later times,
suppression is most severe in the later forming MW-like galaxies.

Finally all distributions contain an excess of [Fe/H] $\sim -3$ stars,
corresponding to unpolluted (Pop III) galaxies.  This discrepancy most
likely relates to our simple treatment of the star-formation within
dwarfs, which fails to account for what is undoubtedly a much altered
process under primordial conditions (e.g. Haiman, Thoul, \& Loeb 1996;
Nakamura \& Umemura 1998).  Note that this discrepancy is
qualitatively in agreement with Hernandez and Ferrara (2000),
although, in our case the distribution of low-metallicity stars is
somewhat complicated by our imposing at all redshifts the finite
minimum mass scale set by photo-ionization.

\section{Conclusions}

We have considered the formation of the Galactic Halo in
the context of hierarchical structure formation models.  While
extensive theoretical and observational evidence suggests that the
Halo was formed primarily from accreted dwarfs, a detailed
understanding of the relationship between this process and the
metallicity distribution of the Halo has yet to be developed.

Using a simple semi-analytic model, we have investigated the
qualitative features that can be expected in a complete picture.
Galaxy outflows enrich the IGM slowly over time, causing the stars in
early-forming galaxies to be more metal poor than those in more
massive later-forming objects.  This results in a ``metallicity
floor'' in the distribution of Galactic disk stars at a level roughly
coincident with the high end of the distribution of halo stars, as
observed.  While previous models of hierarchical formation are at odds
with the observed lack of local group dwarf galaxies, accounting for
``baryonic stripping'' allows us to construct realistic metallicity
distributions while reducing the number of low-luminosity objects.

Our models cannot claim to be complete, but rather point to the
features that can be expected in a fuller
hydrodynamical treatment.  The sensitivity of our results to the
strength of the outflow winds and the degree of baryonic stripping
suggests that a detailed understanding will require more careful
numerical modeling of structure formation and outflows which must be
consistent with the degree of halo velocity and metallicity
substructure inherent to accretion. It is clear that only when we
have understood the enrichment history of the Universe and its
relation to structure formation can we hope to develop a completely
successful model for the metal distribution of stars in our Galaxy.

\acknowledgments 

We thank Andrea Ferrara and the referee, Xavier Hernandez, for helpful
comments and discussions.

\fontsize{9}{11pt}\selectfont

\newpage

\begin{figure}
\centerline{ 
\psfig{file=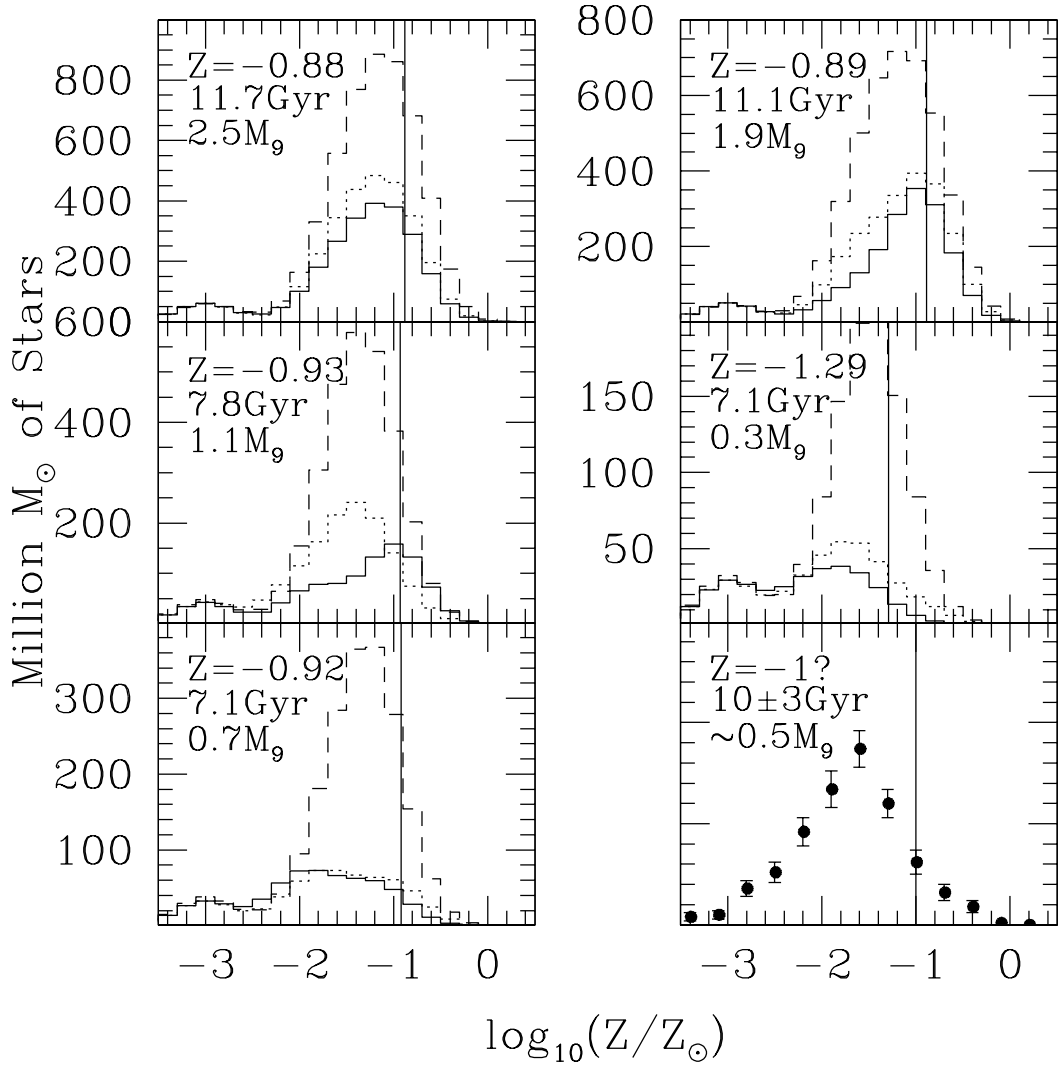,width=5.8in}}
\caption{Distribution of metallicities in the halos  of
MW-like objects.  The solid lines are the results of the fiducial
$\epsilon_{\rm wind} = 0.1$ outflow model, the dotted lines are from
the low-energy outflows simulation $\epsilon_{\rm wind} = 0.05$,
and the dashed lines are the result of including halos which
fail to pass the baryonic stripping criteria, Eq.\ (\protect\ref{eq:strip}),
in the $\epsilon_{\rm wind} = 0.1$ model.
Each panel is labeled with the formation time, halo mass in units of 
$10^9 M_\odot$, and initial metallicity of
the disk component in the fiducial model, which is also labeled by the
vertical lines.  In the lower right panel
we plot the metallicity distribution of the sample of MW Halo stars
studied by Ryan and Norris (1991), arbitrarily normalized.}
\label{fig:hs}
\end{figure}


\begin{references}
\reference{} Abell, G. O. 1958, A\&AS, 50 ,241
\reference{} Argast, D., Samland, M., Gerhard, O. E., \& Thielemann F.-K. 
	2000, A\&A, 356, 873
\reference{} Balland, C., Silk, J., \& Schaeffer, R. 1998, ApJ, 497, 541
\reference{} Bekki, K.  \& Chiba, C. 2000, ApJ, 534, 89L
\reference{} Binney, J. \& Merrifield M. 1998, Galactic Astronomy,
	(Princeton: Princeton Univ. Press)
\reference{} Carney, B. W., Laird, J. B., Latham, D. W., \& Aguilar, L. A.
	1996, AJ, 112, 668	
\reference{}Cole, S., Arag\'on-Salamanca, A., Frenk, C. S., Navarro, J. F., \&
	Zepf, S. E. 1994, MNRAS, 271, 281
\reference{} C\^ote, P., Marzke, R. O., West, M. J., \& Minniti, D. 2000,
	ApJ, 533, 869
\reference{} Dehnen, W. \& Binney, J. 1998, MNRAS, 294, 429
\reference{} Gilmore, G. \& Wyse, R. F. G. 1998, ApJ, 116, 748
\reference{} Haiman, Z., Thoul, A. A., \& Loeb, A. 1996, ApJ, 467, 522
\reference{}Helmi, A. \& White, S. D. M. 1999, MNRAS, 307, 495
\reference{} Hernandez, X. \& Ferrara, A. 2000, MNRAS, submitted
	(astro-ph/0010024)
\reference{} Hernquist, L. 1993, ApJ, 548
\reference{} Ibata, R., Gilmore, G. \& Irwin, M. 1994, Nature, 370, 194
\reference{}Ibata, R, Lewis, G. F., Irwin, M., Totten, E., \&  Quinn, T. 2000,
    ApJ, in press (astro-ph/0004011)
\reference{} Kauffman, G., White, S. D. M., \& Guiderdoni, B. 1993, MNRAS, 264, 201
\reference{} Klypin, A., Kravtsov, A. V., Valenzuela, O., \& Prada,
	F. 1999, ApJ, 522, 82
\reference{} Knox, R. A., Hawkins, M. R. S., \& Hambly, N. C. 1999, MNRAS,
	306, 736
\reference{} Kulessa, A. S. \& Lynden-Bell, D. 1992, MNRAS, 255, 105
\reference{} Larson, R. B. 1974, MNRAS, 166, 585
\reference{} Leggit, S. K., Ruiz, M. T., Bergeron, P. 1998, ApJ, 497, 294
\reference{} Moore, B., Ghinga, F., Governato, G., Lake, T., Quinn,
	T., Stadel, J., \& Tozzi, P. 1999, ApJ, L19
\reference{} Nakamura, F. \& Umemura, M. 1998, ApJ, 515, 239
\reference{} Oey, S. 2000, ApJ, 542, L25
\reference{} Ostriker, J. B., \& McKee, C. F. 1988, Rev. Mod. Phys., 
	60, 1 
\reference{} Ostriker, J.B., \& Thuan, T.X. 1975, ApJ, 202, 353
\reference{} Pagel, B. E J.  1989, RevMAA 18, 153
\reference{} Postman, M. \& Geller, M. J. 1984, ApJ, 281, 95
\reference{}  Preston, G. W, Beers, T. C., \& Shectman, S. A.
	1994, AJ, 108, 538
\reference{} Rohlfs, K. \& Kreitschmann, J. 1988, A\&A, 318, 416
\reference{} Ryan, S. G. \& Norris, J. N. 1991, AJ, 101, 1865
\reference{} Scalo, J. N. 1986, Fundam. Cosmic Phys., 11, 1
\reference{} Scannapieco, E., \& Broadhurst, T. 2001, ApJ, 548, ?? (SB)
\reference{} Scannapieco, E., Ferrara, A., \& Broadhurst, T. 2000,
	 ApJ, 536, L11
\reference{} Scannapieco, E., Thacker, R. J., \& Davis 2000, ApJ, submitted
	(astro-ph/0011258).
\reference{} Schmidt, M. 1959, ApJ, 129, 243
\reference{} Schmidt, M. 1963, ApJ, 137, 758
\reference{} Searle, L. \& Zinn, R. 1978, ApJ, 225, 357
\reference{} Sheth, R. K., Mo, H. J., \& Tormen, G. 1999, MNRAS, submitted,
	(astro-ph/990724)
\reference{} Steinmetz, M. \& Navarro, J. 1999, ApJ, 513, 555
\reference{} Somerville, R. S. 1997, Doctoral Thesis, Univ. of California,
	Santa Cruz
\reference{} Tegmark, M., Silk, J., \& Evrard, A. 1993, ApJ, 417, 54 
\reference{} Truran, J. W. \& Cameron, A. G. W. 1971, Ap\&SS, 14, 179
\reference{} Tsujimoto, T., Shigeyama, T., \& Yoshii, Y. 1999, ApJ, 519, L63
\reference{} White, S. D. M. \& Frenk, C. S. 1991, ApJ, 379, 52
\reference{} Winget D. E. et al.\ 1987, ApJ, 315, L77
\reference{} van den Bergh, S. 1962, AJ, 67, 486


\end{references}
\end{document}